\documentclass[twocolumn,usenatbib]{aa}
\usepackage{graphicx}
\usepackage{txfonts}
\begin{document}

\def\fsec{\hbox{$.\!\!^{s}$}}
\def\fsec{\hbox{$.\!\!^{s}$}}
\def\farcs{\hbox{$.\!\!^{\prime\prime}$}}
\def\hi{\hbox{H\,{\sc i}}}
\def\kms      {\ifmmode{\rm km\,s}^{-1} \else km\,s$^{-1}$\fi}
\def\mujybm{\ifmmode{\rm \mu Jy}\,{\rm beam}^{-1}\else${\rm \mu}$Jy\,beam$^{-1}$\fi}
\def\ltsim{\ifmmode\stackrel{<}{_{\sim}}\else$\stackrel{<}{_{\sim}}$\fi}
\def\gtsim{\ifmmode\stackrel{>}{_{\sim}}\else$\stackrel{>}{_{\sim}}$\fi}
\def\fr{\ifmmode\mathcal R_{1\rightarrow0}\,\else${\mathcal
R}_{1\rightarrow0}$\,\fi}


%
   \title{Sub-arcsecond imaging of the radio continuum and neutral
hydrogen in the Medusa merger}
\authorrunning{{\sl Beswick et al.}}

\titlerunning{H{\sc i} absorption in the Medusa Merger}

   \author{R.\,J.\,Beswick
          \inst{1}
          \and
          S.\,Aalto 
	  \inst{2}
	  \and 
	  A.\,Pedlar
	  \inst{1}
	  \and
	  S. H{\"u}ttemeister
	  \inst{3}       
  }

   \offprints{{Robert.Beswick@manchester.ac.uk (RJB),
susanne@oso.chalmers.se (SA)}}

   \institute{The University of Manchester, Jodrell Bank Observatory, Macclesfield, Cheshire SK11~9DL, UK
         \and
             Onsala Rymdobservatorium, Chalmers Tekniska H{\"o}gskola, 43992 Onsala, Sweden.
	\and
Astronomisches Institut Ruhr Universit{\"a}t Bochum, Universit{\"a}t{\ss}tr, 150, D-44780 Bochum, Germany}

   \date{received; accepted; original form 10th February 2005}

   \abstract{We present sub-arcsecond, Multi-Element Radio Linked
Interferometer (MERLIN) observations of the decimetre radio continuum structure and neutral hydrogen (H{\sc i}) absorption
from the nuclear region of the starburst galaxy NGC\,4194 (the
Medusa Merger). The continuum structure of the central kiloparsec of
the Medusa has been imaged, revealing a pair of compact radio
components surrounded by more diffuse, weak radio emission. Using the
constraints provided by these observations and those within the
literature we conclude that the majority of this radio emission is
related to the ongoing star-formation in this merger system. 

\hskip 0.4cm With these observations we also trace deep H{\sc i}
absorption across the detected radio continuum structure. The
absorbing H{\sc i} gas structure exhibits large variations in column densities. The largest column densities are
found toward the south of the nuclear radio continuum, co-spatial with
both a nuclear dust lane and peaks in $^{12}$CO\,(1$\rightarrow$0)
emission. The dynamics of the H{\sc i} absorption, which are
consistent with lower resolution $^{12}$CO emission observations,
trace a shallow north-south velocity gradient of
$\sim$320\,km\,s$^{-1}$\,kpc$^{-1}$. This gradient is interpreted as
part of a rotating gas structure within the nuclear region. The H{\sc
i} and CO velocity structure, in conjunction with the observed gas
column densities and
distribution, is further discussed in the context of the fuelling and
gas physics of the ongoing starburst within the centre of this merger.

   }

   \maketitle

%
	
\section{Introduction}

The Medusa merger (NGC\,4194) is a nearby (D=39\,Mpc) star-forming
galaxy of intermediate infrared luminosity (L$_{\rm
FIR}=8.5\times10^{10}$\,L$_{\odot}$). As such the Medusa merger falls
into a luminosity class, considerably lower than the well-studied
ultra-luminous infrared galaxies (ULIRGs) but it still appears to show
areas of intense starburst activity (Armus, Heckman \& Miley 1990;
Prestwich, Joseph \& Wright 1994). The optical morphology of the
Medusa merger, as its name suggests, traces clumpy and extended
features which stretch $\sim1\arcmin$ to the north of the main optical
nucleus. These `hair-like' extensions, indicative of tidal tail
features resultant from the galaxy's ongoing merging event, contain
significant volumes of molecular gas (Aalto, H{\"u}ttemeister \&
Polatidis 2001) although no indications of ongoing star formation in
the tidal tail have, as yet, been detected (Armus et al. 1990). The
majority of star formation in the Medusa merger is concentrated in the
extended nuclear starburst (Armus et al. 1990). 

A detailed study of the molecular gas (CO J=1$\rightarrow$0) composition and kinematics
in NGC\,4194 has been previously made at arcsec angular resolutions at
the Owens Valley Radio Observatory (OVRO) by Aalto \&
H{\"u}ttemeister (2000, hereafter AH00).  This investigation revealed
that the $^{12}$CO emission is extended over an area of 25\arcsec
(4.7\,kpc) covering the central and north-eastern part of the optical
galaxy. The CO emission also traces the two prominent dust lanes that
cross the central region and extends into the northern tidal tail. The
majority of the CO ($\sim$70$\%$) is found with the central 2\,kpc of
the galaxy, with 15$\%$ of this gas residing in a compact region
1\farcs5 south of the radio nucleus.    

We have used the UK's Multi-Element Radio Linked Interferometric
Network (MERLIN) to observe both the 1.4\,GHz radio continuum and
neutral hydrogen (H{\sc i}) absorption structure of the central few
hundred parsecs of NGC\,4194 at sub-arcsec angular resolutions. This
constitutes the highest angular resolution study made to date of both
the radio continuum and neutral gas structure in this merger
galaxy. In this paper we present the results of these observations and
place them in context with other observations of the radio continuum,
and  neutral and molecular gas in NGC\,4194. The primary goal of this
work is to provide high linear resolution information about the
neutral gas within the core region of this star-forming galaxy,
enabling the composition and dynamics of the central region to be
investigated. In conjunction with lower resolution data ({\sl e.g.}
AH00;  Aalto et al. 2001) these observations can be used to more
fully chart the gas dynamics and fuelling of this intermediate
luminosity starburst galaxy. 

This paper is organised as follows. The observations and the data
reduction processes undertaken are described in detail in the
next section. This is followed by the presentation of the radio
continuum and H{\sc i} absorption results. In the subsequent sections
we will discuss, in detail, these observational results, initially
drawing conclusions regarding the neutral hydrogen and radio continuum
structure of the central 0.5 kiloparsec region of the Medusa and then
by placing these results in the context of other observations of this
source's gas content.

Throughout this paper we will assume a distance of 39\,Mpc to
NGC\,4194 (H$_0$$=$75\,km\,s$^{-1}$Mpc$^{-1}$), implying that at the
distance of the source 189\,pc subtends an angle of 1 second of arc.

\section{Observations and Data Reduction} 
\begin{figure*}
   \centering
    \includegraphics[angle=0,width=16cm]{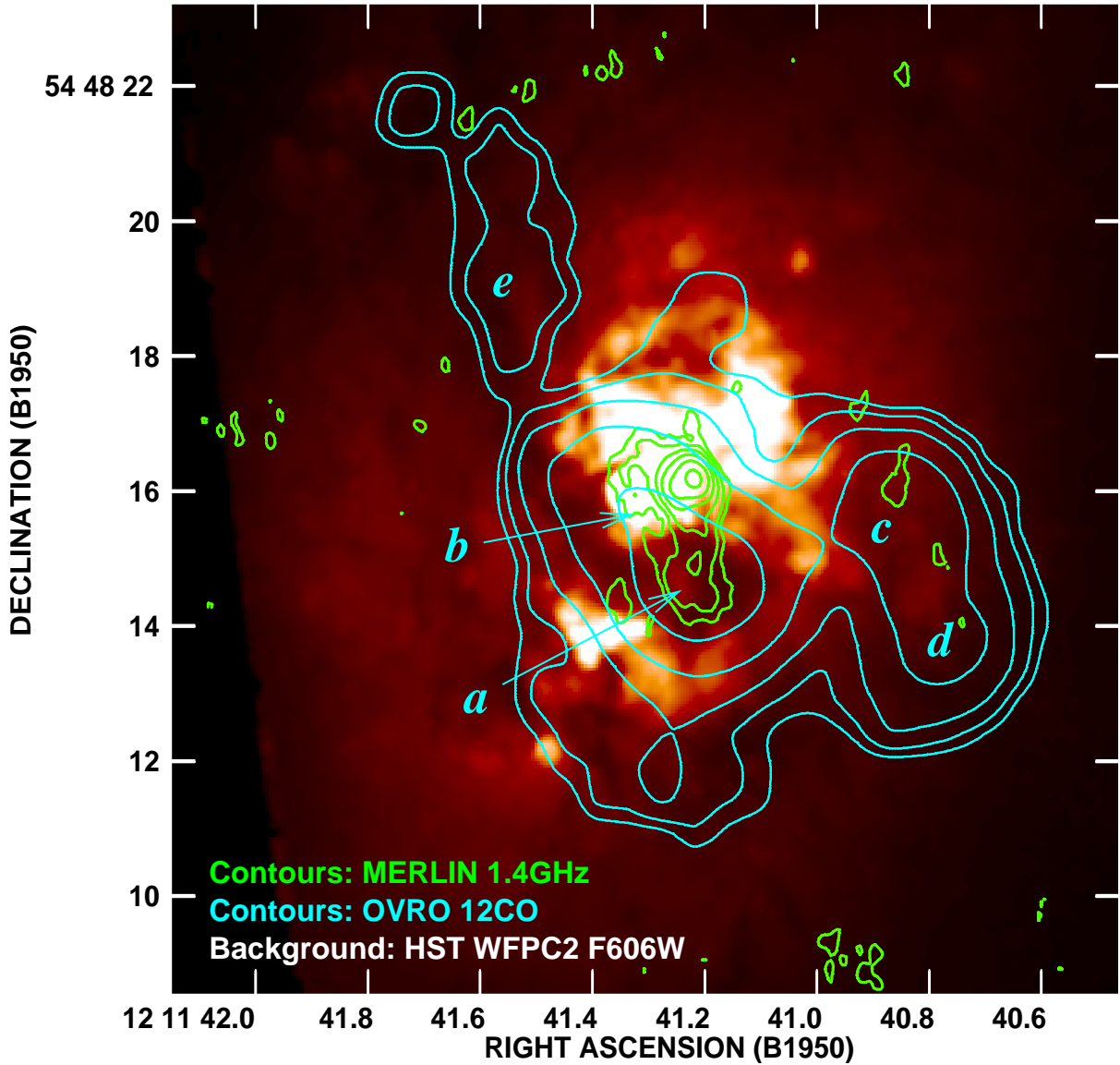}
      \caption{Overlay of {\sl uv}-tapered (800\,k$\lambda$) 1.4\,GHz
radio continuum emission in green contours and OVRO $^{12}$CO (AH00)
in  pale blue contours over a F606W WFPC2 {\sl HST}
image. The contour levels on the radio continuum image are $-$1, 1, 2,
4, 8, 16 and 32 times 0.3\,mJy\,bm$^{-1}$. The $^{12}$CO contours are
1, 2, 4, 8, 16 and 32 times 0.5\,Jy\,beam$^{-1}$\,km\,s$^{-1}$. The false-colour scale of the
optical image is linear and arbitrary. The radio and CO images have synthesised beams of 0\farcs446$\times$0\farcs320 and
2\farcs06$\times$1\farcs70 respectively. The astrometrical alignment of
the {\it HST} image with the other two images has an rms error of $\sim$100\,mas and is described
in section 2.2. Labels {\sl a, b, c, d} and {\sl e} indicate the
$^{12}$CO peaks as observed by AH00.} 
             
         \label{Fig1}
\end{figure*}

\subsection{MERLIN}

NGC\,4194 was observed using the MERLIN array,
including the 76-m Lovell Telescope (Thomasson 1986), on 2003 March 15
at 1408.66\,MHz. NGC\,4194 was observed for $\sim$18\,h interspersed with
regular observations of the nearby phase calibration source
1205+545. Throughout the observation dual hands of circular
polarisation were recorded over an 8-MHz band which were correlated
into 64 spectral channels, each of bandwidth 125\,kHz providing a
velocity resolution of $\sim$26\,km\,s$^{-1}$\,channel$^{-1}$, centred upon
the rest velocity of NGC\,4194, 2560\,km\,s$^{-1}$. A single
observation of OQ208 was used to obtain passband solutions along with
a 30\,min integration on 3C\,286, which was assumed to have a flux
density of 14.794\,Jy at 1408\,MHz (Baars et al. 1977) and was
subsequently used to calibrate the flux density scale for all sources.

Initial editing and calibration of these NGC\,4194 data were carried out using
local MERLIN software at Jodrell Bank Observatory, prior to these data
being read into the Astronomical Image Processing Software ({\sc
aips}). Within {\sc aips} further editing and bandpass calibration
were applied to the data set. Phase corrections derived from the phase
calibrator source 1205+545 were also applied to these data, along with
further phase corrections derived from self-calibration upon the
continuum of NGC\,4194. The calibrated {\sl uv} data set was
Fourier-transformed to produce spectral line data cubes with various
tapering and data weighting schemes used in order to produce images
that were sensitive to diffuse large scale radio emission and high
resolution images of the compact components. The line-free channels
were combined to produce continuum images which were subsequently
subtracted from the spectral-line cubes. These continuum images along
with the continuum subtracted spectral-line cubes were deconvolved
using a {\sc clean} based algorithm to remove the instrumental
response (H{\"o}gbom 1974). The cleaned continuum images and
continuum-subtracted cubes were then recombined and further analysis of
these data were undertaken using standard spectral line tasks within
{\sc aips}.

These MERLIN data have a shortest projected baseline of $\sim$13\,km
between the Tabley and Jodrell Bank telescopes and hence these data
are not sensitive to extended components greater than a few
arcsecs. The consequences of this data-set including few low-order
spacings is a hole in the {\sl uv} coverage toward the centre of the
{\sl uv} plane. This feature of sparsely filled interferometric
arrays, such as MERLIN, causes an under-sampling of the low-order
spacings and hence an insensitivity to regions of smooth, extended
flux. This effect can result in a bowl-shaped reduction in the flux
density toward extended sources (see, for
example Wills, Pedlar \& Muxlow 1998).  This not only reduces the flux
density of radio continuum components derived from images but also
causes the calculated H{\sc i} absorption opacities to be over
estimated due to the reduction in the value of the extended continuum
emission. The total flux measured from these observations is 28\,mJy
compared to a lower resolution (beam size of 45\arcsec) VLA 1.4\,GHz flux of 91\,mJy (Condon et
al. 1990). The distribution of the radio emission detected in
this VLA image covers an area is approximately 4 times larger
than the region shown in Fig.\,1. This implies that these observations are spatially filtering out
approximately two thirds of the 1.4\,GHz flux. If it is assumed that
this missing flux is spread over a Gaussian with an area matching the
component detected by Condon et al., the maximum correction to the
observed MERLIN map that is derived is 0.05\,mJy at any single point
in the image. In the case of the observations presented in this paper
we have further tested for the significance of these missing low-order
spacings by taking slices across the images in several
directions. Although some evidence for a slight reduction in the flux
density of the images produced was found, it has not been directly
corrected for in the images and spectra present since the correction
required was found to be significantly smaller than the channel based
noise levels in these data.  

\begin{figure}
   \centering
   \includegraphics[angle=0,width=8cm]{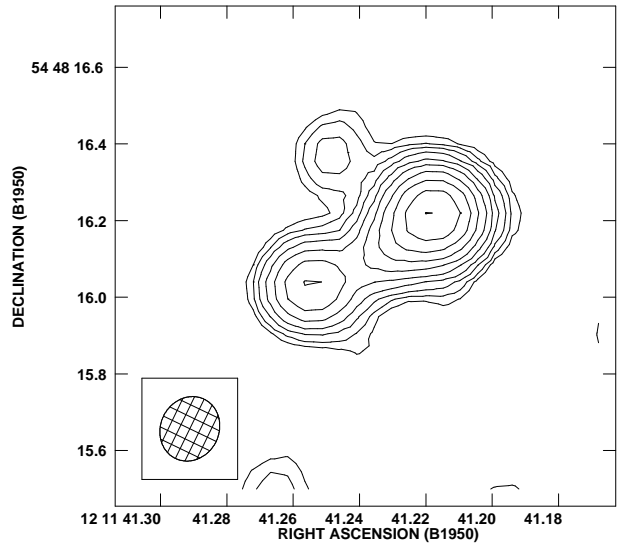}
      \caption{Uniform weighted MERLIN 1.4\,GHz radio continuum image
of the core of the Medusa merger with a synthesised restoring beam of
0\farcs17 $\times$ 0\farcs15. Contour levels start at
0.35\,mJy\,bm$^{-1}$ and increase by factors of $\sqrt{2}$.}
             
         \label{Fig2}
\end{figure}

\subsection{Hubble Space Telescope (HST)}

In order to afford a good comparison between the sub-arcsec radio
observations presented in this paper and the optical emission from this galaxy
archival {\sl HST} observations have been acquired. These NGC\,4194
data (U2E66Y01T) consisted of single 600\,s integration using a
broad-band F606W filter (see Fig.\,1). Initially these data were reduced using both
{\sc stsdas} commands within {\sc iraf} and {\sc starlink} software
before being exported as a {\sc fits} file. The procedure of applying
accurate astrometry to the {\sl HST} field was achieved by assigning known
coordinates, obtained from the US Naval Observatory star catalogue, to
a number of stars spread throughout the WFPC-2 field. This procedure
resulted in a linear transformation and rotation being applied to the
{\sl HST} image in-order to minimise the residual position errors of
the known star positions. No shear was applied to the image during
this process. It is estimated that the absolute positional accuracy of
the aligned MERLIN and {\sl HST} images is $<$0.1\,arcsec.
  
  \begin{table*}
      \caption[]{The 1.4\, GHz radio continuum properties of the core
components.$^{\mathrm{a}}$} \label{tab1}
     $$
         \begin{array}{p{0.15\linewidth}p{0.15\linewidth}p{0.12\linewidth}p{0.11\linewidth}p{0.14\linewidth}p{0.14\linewidth}p{0.12\linewidth}}
            \hline
            \noalign{\smallskip}
            R.A.\,\,(B1950) 12$^{\rm h}11^{\rm m}$&Dec.\,\,(B1950)
54$\degr 48\arcmin$& Peak flux\,density (mJy\,bm$^{-1}$)&Integrated
flux\,density (mJy)& Angular size&Linear size (pc)& P.A. \\
            \noalign{\smallskip}                 
            \hline
            \noalign{\smallskip}
               41.$\!\!^{\rm
s}$22&16\farcs210&8.20&12.6&0\farcs140$\times$0\farcs093&27.8$\times$17.6&97.$\!\!\degr$9   \\
	       41.$\!\!^{\rm s}$25&16\farcs038&2.84&5.11&0\farcs187$\times$0\farcs097&35.3$\times$18.3&106.$\!\!\degr$6\\
	      
            \noalign{\smallskip}
            \hline 
         \end{array}
     $$
\begin{list}{}{}
\item[$^{\mathrm{a}}$] Listed properties are for the two compact
nuclear components. Flux densities and positions have been derived
from Gaussian fits to our uniformly weighted image (Fig\,2). 
\end{list}
\end{table*}

\section{Results}
\subsection{The 1.4\,GHz radio continuum}

A {\sl uv}-tapered (800\,k$\lambda$) 1.4\,GHz image of the central
region of NGC\,4194, derived from the line-free channels, is shown as
contours in Fig.\,1, overlayed upon a false-colour archival WFPC2
{\sl HST} image.  A {\sl uv}-taper was applied to these data in order to
produce images which are as sensitive as possible to the weak and
diffuse extended radio emission. A measured noise level of
 0.1\,mJy\,bm$^{-1}$ has been determined for Fig.\,1 and the image
has an angular resolution of 0\farcs45$\times$0\farcs32. Figure 2
shows a high angular resolution, uniformly weighted, contour image of
the core region of the galaxy. This figure has an angular resolution
of 0\farcs17$\times$0\farcs15 and a noise level of
$\sim$0.1\,mJy\,bm$^{-1}$.  Table\,1 lists the 1.4\,GHz flux
densities and Gaussian fitted sizes of the two, bright, compact components seen
in the uniformly weighted image (Fig.\,2). The radio structure of the
central few hundred parsecs of NGC\,4194 is dominated by a pair of
compact radio components (Fig.\,2). These two components are separated
by $\sim$0\farcs35 ($\sim$65\,pc) along a PA of 120$\pm$5\degr. In
the {\sl uv}-tapered image shown in Fig.\,1, and enlarged as the
central image of Fig.\,3,  it can be seen that these two compact radio
components are only partially resolved and are embedded in a halo of
diffuse radio continuum which extends in a southerly direction.   

\begin{figure*}[th!]
   \centering
   \includegraphics[angle=0,width=18cm]{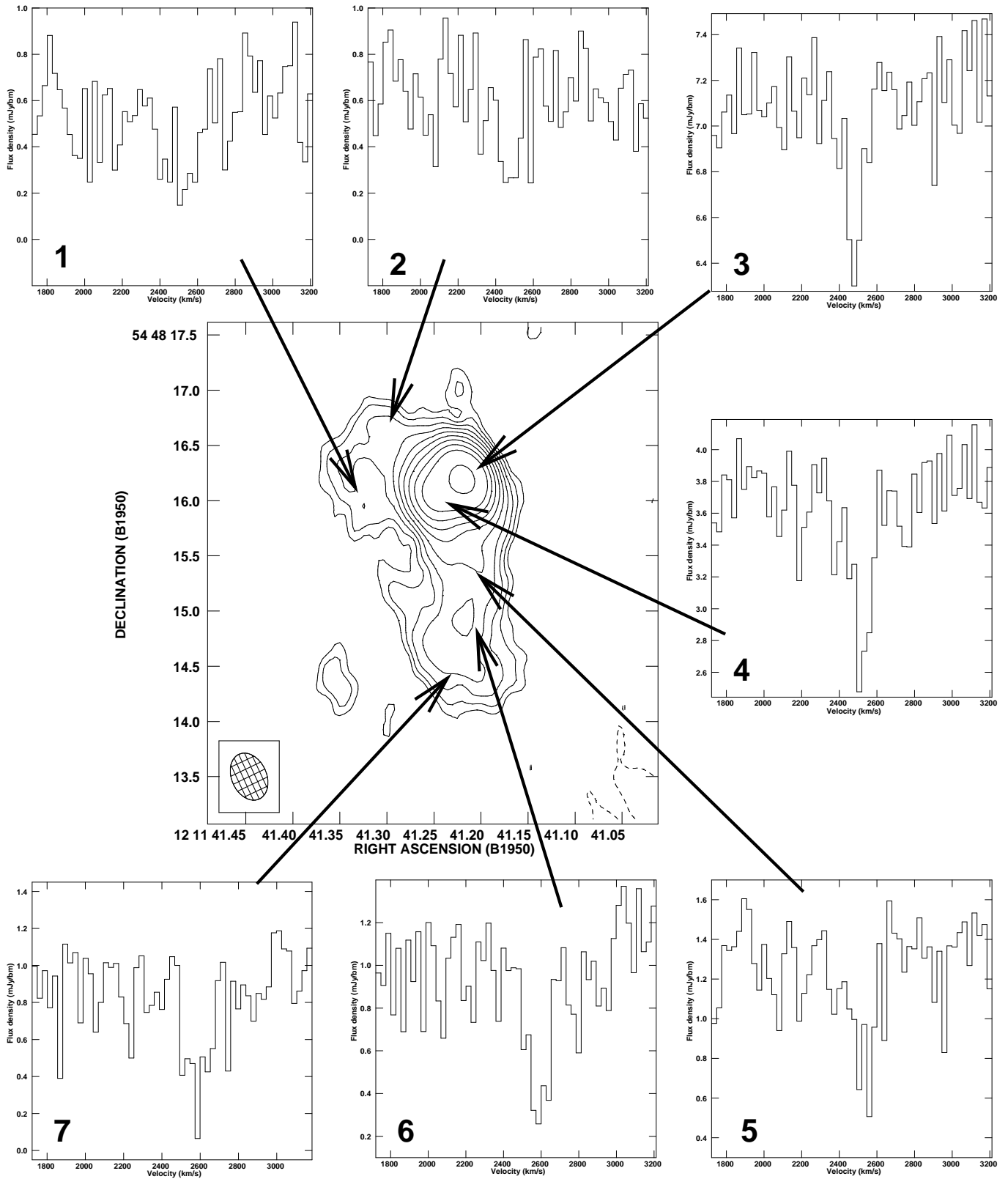}
    \caption{Montage of 800-k$\lambda$ {\sl uv}-tapered radio
continuum structure and H{\sc i} absorption spectra. Contour levels of
the central image start at 0.3\,mJy\,bm$^{-1}$ and increase by factors of $\sqrt{2}$ with an
angular resolution of 0\farcs45 $\times$ 0\farcs32.}
         \label{Fig3}
\end{figure*}

 \begin{figure*}
   \centering
\setlength{\unitlength}{1mm}
    \begin{picture}(80,75)
\put(0,0){\includegraphics{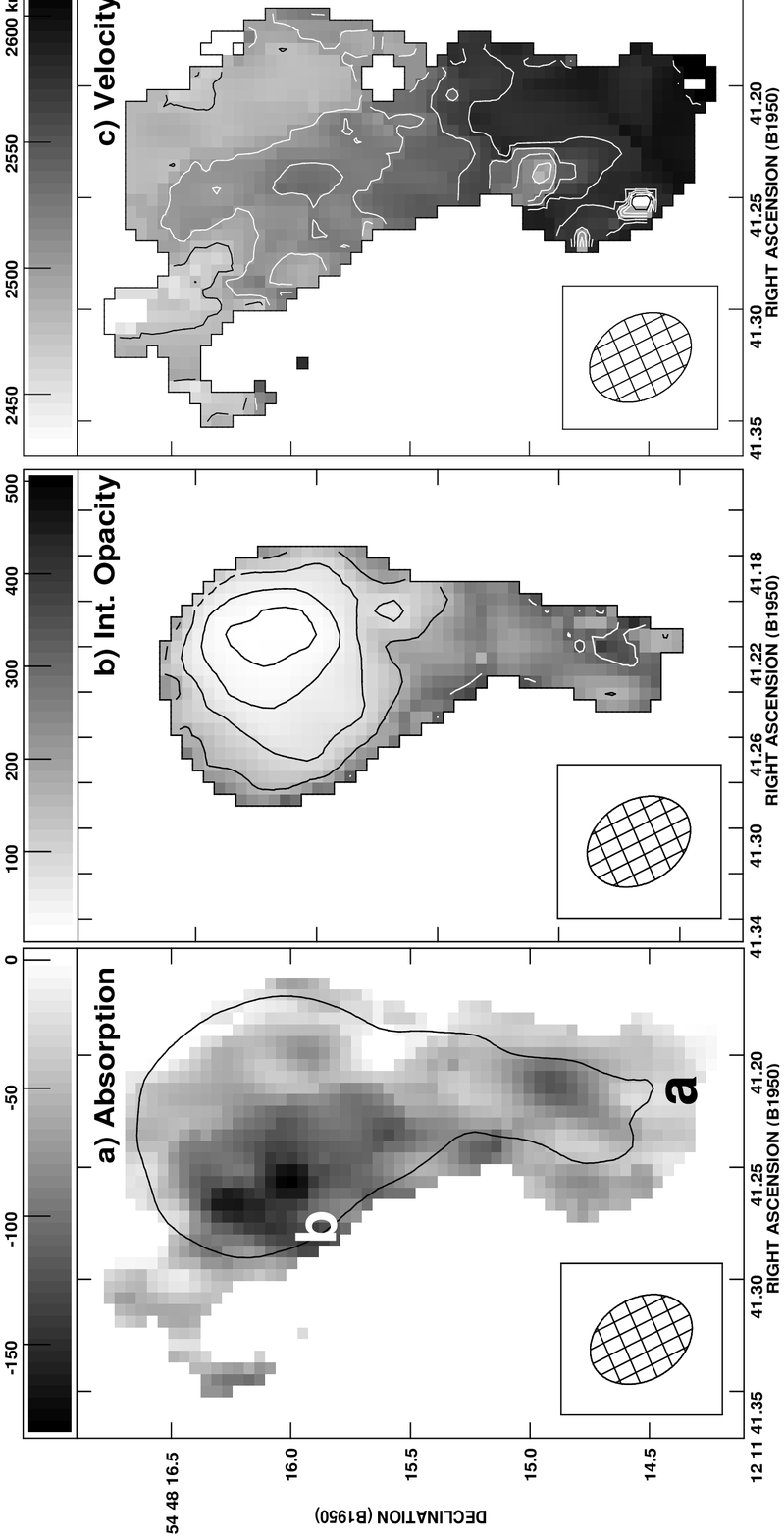}}
\end{picture}
\vskip 1cm

      \caption{ {\bf a)} The (leftmost) image shows the integrated
H{\sc i}
absorption signal across the
centre of the Medusa merger. This has been derived from the continuum
subtracted spectral line cube. The grey-scale flux density range is
0$\rightarrow-$185\,mJy\,bm$^{-1}$\, km\,s$^{-1}$.   Note the absorption shown is
dependent on the strength of the continuum emission. The two symbols
(a and b) shown on the left-most image represent two of the CO
emission peaks from AH00. A single
contour 1\,mJy\,bm$^{-1}$ radio continuum contour is overlayed. Areas
external to this contour have the lowest levels of radio continuum and
have been excluded in the derivation of Fig\,4b. {\bf b)} The (central) image shows the 
integrated optical depth ($\int \tau$dV) across the same region as
{\bf a}. The grey-scale range shown is 15$\rightarrow$500 and the contour
levels are 10, 20, 40, 80, 160, 320 and 640 times 100. It should be
noted that errors in this image increase in areas where the radio
continuum flux density is at a minimum. This effect results in some
erroneous values toward the edges of the image, in order
to remove the most severe of these effects this image has been blanked
in areas where the radio continuum is less than
1\,mJy\,beam$^{-1}$. {\bf c)} The (right-hand) map represents the H{\sc
i} absorption velocity field over areas where significant radio continuum and
absorption line signal were detected. The grey scale range of the
image is 2430$\rightarrow$2610\,km\,s$^{-1}$, the overlayed contours
have increments of 25\,km\,s$^{-1}$ between
2475$\rightarrow$2650\,km\,s$^{-1}$. }
            \label{Fig4}

\end{figure*}

\subsection{Neutral hydrogen absorption}

As can be seen in Fig.\,3, extensive H{\sc i}
absorption is detected against the majority of the background radio
continuum in the {\sl uv}-tapered MERLIN image. Similarly
H{\sc i} absorption is also found against the core region in
spectral line cubes made with uniform weighting. The H{\sc i}
absorption properties derived from the {\sl uv}-tapered spectral line
cube are listed in Table\,2. These properties include the central
positions of the spectra shown in Fig.\,3, in addition to the Gaussian
fitted velocities and widths of the absorption lines. Table\,2 also
lists derived properties of these absorption lines, such as their peak
opacities and column densities of H{\sc i} along these lines of
sight. The column densities listed have been calculated assuming
that N$_{\rm H}=1.823\times 10^{18}{\rm T_{spin}}\int{\tau}{\rm dV}$.
 Although T$_{\rm spin}$ can be reasonably assumed to be 100\,K (Maloney,
Hollenbach \& Tielens 1996) it should be noted that the value of
T$_{\rm spin}$ is dependent upon the physical conditions of the gas
and will therefore vary significantly especially in the extreme
environment of a nuclear starburst. These considerations will be
discussed in more detail in a latter section.     

In Fig.\,4 we present  three grey-scale images of the centre of the
Medusa presenting the H{\sc i} absorption detected against the
radio continuum. The left-most (Fig.\,4{\bf a}) of these panels shows a
grey-scale map of the H{\sc i} absorption toward areas with
significant radio continuum background. As can be seen in this image the line
signal varies, with the absorption deepest toward the
eastern edge of the brightest radio component. The map of
integrated absorption shown in Fig.\,4{\bf a} does not represent the true
H{\sc i} column density since it is dependent upon the background
continuum strength. In the central panel (Fig.\,4{\bf b}) the
distribution of the integrated H{\sc i} opacity is shown. This panel
shows the distribution of neutral hydrogen and is directly proportional
to the H{\sc i} column density. It should, however, be noted that the
process of deriving maps of optical depth involves the division of
spectral line data-cubes with their associated radio continuum images, resulting in the opacity images produced being susceptible to
errors in areas where the radio continuum is small ({\sl i.e.} toward
the edges of the image). Measures to minimise this effect have been
made, although some unphysical values in a few
pixels toward the edges of the image may remain. However considering
this factor the variations in opacity seen are significantly larger
than the errors in the image. Plus the values and features seen in this image are consistent with
those observed and calculated from the absorption spectra (see
Fig.\,3) and absorption maps
(Fig.\,4{\bf a}). The H{\sc i} integrated opacity shows an increase in
the
column densities south of the radio continuum peak which is consistent with the
positions of the peaks of the CO emission observed by AH00 (labelled
in Fig.\,1 \& 4{\bf a}). 

Figure\,4{\bf c} shows that the velocity structure of the H{\sc i} follows a
distinct north to south velocity gradient on the largest scales along
with an N-W to S-E velocity gradient across the two bright, compact
radio components. Fig.\,5 shows a position (Declination)
verses velocity contour plot of the absorption averaged over the entire
R. A. extent of the radio continuum source.  This figure complements
the two dimensional velocity distribution shown in Fig.\,4{\bf c} by
showing in detail the north to south velocity gradient in the
source.  As can be seen from these figures the velocity structure illustrated
in Fig.\,4{\bf c} traces a gradient of $\sim$60\,km\,s$^{-1}$arcsec$^{-1}$.

\section{Discussion}
\subsection{High resolution radio continuum structure of the centre of the
Medusa}

In the highest angular resolution (0\farcs17$\times$0\farcs15) image
produced using these MERLIN data (Fig.\,2) the majority of the radio
structure of the Medusa galaxy is resolved away leaving a pair of
compact nuclear components (see Table\,1 \& Fig.\,2) with a small,
weak extension toward the north-east of the brightest component. This
double structure is reminiscent of the core regions of many low
powered Seyfert galaxies ({\sl e.g.} Thean et al. 2000).  However, the
lower angular resolution {\sl uv}-tapered images show these
components to be sitting in an
extensive, diffuse, radio emission toward the south of these compact
components (see Figs.\,1 and 3).  These {\sl uv}-tapered images
(Fig.\,1 and 3) are consistent with other studies of
this source previously made at higher frequencies using the Very Large
Array (VLA) (e.g. Ulvestad, Wilson \& Sramek 1981).  Fig.\,3 is still
insensitive to many of the weak, larger scale features observed by
Ulvestad et al. (1981). More recent multi-frequency (5 to 15\,GHz) VLA observations
(S.~Neff, private comm.; Neff, Campion \&
Ulvestad 2001) show that in
addition to the compact components and diffuse emission
observed in these MERLIN 1.4\,GHz data more
extended emission is present. In particular, in deep 5\,GHz VLA
images (S.~Neff, private comm.), an
`arm-like' extension arcing from the north of the nuclear
region and ending approximately 4\,arcsec to the west of this region
in a strong compact component, is observed. This feature is also seen in
VLA images published by Ulvestad et al. (1981). Although some traces
of this `arm-like' emission can be seen in our {\sl uv}-tapered images (see Fig.\,1) the majority of these features are
not detected by our observations. Neff et al. (2001) also report
$\sim$50 compact sources ($>5\times10^{18}$\,W\,Hz$^{-1}$). Of the brightest of these
compact sources $\sim$50$\%$ nominally have flat spectral indices and
maybe associated with thermal radio emission from H{\sc ii} regions,
whereas the remainder have steeper spectral indices that are
indicative of emission dominated by non-thermal mechanism (such as
supernovae remnants or AGN). Our sub-arcsec 1.4\,GHz observations
are, as mentioned above, insensitive to much of the diffuse extended
radio emission detected with the VLA, but more sensitive to
emission from synchrotron sources than thermal sources;
the combination of these effects and our sensitivity levels accounts
for the differences in observed emission structures.       

\subsection{The neutral hydrogen absorption}

High optical depth ($\tau_{\rm peak}>$0.5) neutral hydrogen absorption
is detected against almost all of the radio continuum emission
detected (Fig.\,1). The absorbing H{\sc i} gas is observed
to show significant variations in opacity across the radio
continuum source. The H{\sc i} column densities observed against the
radio continuum emission are consistently higher
toward the southern half of the nuclear radio continuum.  These
variations can be seen in both the maps of the integrated H{\sc i}
absorption and opacity (Fig.\,4{\bf a},~{\bf b}) as well as in the
individual spectra (Fig.\,3 and Table\,2). This implies that the H{\sc
i} observed, is not part of a uniform homogeneous screen of neutral
foreground gas. It is likely that the large scale variations in the H{\sc
i} opacities observed are the result of high gas column densities
along the lines of sight of other large scale features, such as the
dust lanes. Whereas smaller, more localised, variations seen in
Fig.\,4{\bf b} (on size scales of order a beam [$\sim$50--60pc]) are likely
to be due to single, or the superposition of several, gas clouds along
the sight line. In Fig.\,1 the distribution of the radio continuum
relative to the WFPC2 R-band image shows a thick dust lane arc that
crosses the southern part of the MERLIN radio continuum in an
approximately east-west direction. The position of this dust lane, which is also co-spatial
with the CO emission (red contours in Fig.\,1; AH00), is coincident with the area of highest
opacity H{\sc i} absorption. It is also reasonable to assume that
the gas, both CO and H{\sc i}, is associated with this dust lane (the
relative distributions of these constituents is discussed in more
detail in Sec. 4.4.1). Hence this implies that the radio continuum
originates from deeper within the source than these components or that
these elements of the neutral ISM and the radio emitting material are
mixed resulting in these observations underestimating the true H{\sc
i} column densities. 

The H{\sc i} velocity field shown in Fig.\,4{\bf c} traces a dominant
but shallow velocity gradient with a north to south direction. This
gradient is also shown in Fig.\,5. In
this averaged position-velocity slice the broad absorption tracks
a velocity gradient of $\sim$60\,km\,s$^{-1}$arcsec$^{-1}$
($\sim$320\,km\,s$^{-1}$kpc$^{-1}$). This velocity gradient extends
over $\sim$2\farcs5 (470\,pc) of the background
radio continuum source. The north-south orientation of the velocity
field is consistent with high resolution CO observations from the same
region (AH00). The details of the velocity field (Fig.\,4{\bf c}) show
that the primary gradient deviates from a north to south direction
against the brighter northern radio continuum components. Hints of this directional change in the
H{\sc i} velocity gradient are visible in the highest angular
resolution ($\sim$1\farcs7) CO data presented by AH00.

   \begin{table*}
      \caption[]{The H{\sc i} absorption properties of the central region
of the Medusa merger.}\label{h1tab}
     $$
         \begin{array}{p{0.13\linewidth}p{0.13\linewidth}p{0.13\linewidth}p{0.13\linewidth}p{0.13\linewidth}p{0.13\linewidth}p{0.13\linewidth}p{0.13\linewidth}}
            \hline
            \noalign{\smallskip}
            Label$^{\rm a}$ &R.A.\,\,(B1950)$^{\rm a}$ 12$^{\rm h}\!11^{\rm m}$&Dec.\,\,(B1950)$^{\rm a}$ 54$\degr\!48\arcmin$& ${\mathrm V}_{\mathrm H}$ (km\,s$^{-1}$)&$\Delta$V\,(km\,s$^{-1}$)  ~($\sigma$)&$\tau_{\mathrm peak}$& N$^{\rm c}\!\!_{\rm H}\,\,\times$10$^{19}\,{\rm T_{\rm spin}}$ (atoms\,cm$^{-2}$) \\

            \noalign{\smallskip}                 
            \hline
            \noalign{\smallskip}
               1$^{\rm b}$&41.$\!\!^{\rm s}$32&16\farcs085&-&-&-&- \\
	       2$^{\rm b}$&41.$\!\!^{\rm s}$31&16\farcs545&-&-&-&- \\
       3&41.$\!\!^{\rm s}$22&16\farcs206&2482&41&0.115  & 0.08  \\
       4&41.$\!\!^{\rm s}$25&16\farcs059&2520&49&0.533  & 2.02\\
       5&41.$\!\!^{\rm s}$23&15\farcs635&2528&58&0.964  & 11.27\\
       6&41.$\!\!^{\rm s}$22&14\farcs915&2587&42&1.350  & 4.82 \\
       7&41.$\!\!^{\rm s}$25&14\farcs600&2586&52&1.425  & 5.89\\
            \noalign{\smallskip}
            \hline 
         \end{array}
     $$
\begin{list}{}{}
\item[$^{\mathrm{a}}$] Labels and positions listed refer to the centre
of the areas over which the spectra in Fig.\,3 are averaged.
\item[$^{\mathrm{b}}$] No limits upon the optical depth are given in
areas where the radio continuum flux density is too low to provide a
physically important limit.
\item[$^{\mathrm{c}}$] Column densities have been calculated assuming
N$_{\rm H}=1.823\times 10^{18}{\rm T_{spin}}\int{\tau}{\rm dV}$.
These are derived from the absorption opacities and line widths over
areas of approximately a beam in size. The value of T$_{\rm spin}$ is
assumed for the purposes of the discussion to be 100\,K, however this
variable will be discussed in more detail in Sec.\,4.4.3.
\end{list}
\end{table*}

\subsection{The source of the radio emission in the Medusa}

Although the Medusa merger is an order of magnitude fainter at
infrared wavelengths than the well studied ULIRG galaxies, such as
Arp\,220, it has in the recent past undergone, and continues to undergo,
significant star-formation activity. This activity has presumably been
triggered by its advanced merger state. However, it is also widely
accepted that gas rich mergers, such as NGC\,4194, can also
host AGN activity in the remnant nuclei of the merged sources
(Genzel et al. 1998). This phenomenon is observed in some composite
starburst-AGN sources ({\sl e.g.} Beswick et al. 2001;
Gallimore \& Beswick 2004; Kewley et al. 2000) and is also evident in
some minor mergers which host more powerful AGN such as radio
galaxies (Heckman et al 1986). As discussed previously the radio continuum
observed from the Medusa merger, at 1.4\,GHz, traces both the areas of
starburst activity, as well the more compact radio components that may
be AGN related. In the rest of this section we will discuss both
the nuclear starburst of the Medusa and its rate of star-formation, as
well as briefly explore the possible evidence that a weak AGN exist in
the core of this galaxy.     

\subsubsection{An AGN component?}

 NGC\,4194 is optically classified as a peculiar, blue compact galaxy
with a strong H{\sc ii} emission line spectra and a blue continuum
indicative of a young stellar population (Liu \& Kennicutt
1995). However, as discussed above the two compact radio components
imaged in the centre of this galaxy show a structure that is similar to that observed in many low luminosity Seyfert
galaxies. These two components are relatively low in luminosity,
5.48$\times10^{20}$ and 2.22$\times$10$^{20}$\,W\,Hz$^{-1}$, but are
compact on MERLIN scales ($<$25\,pc), implying that the strongest of
these components has a brightness temperature in excess of 4.5$\times10^4$\,K. Searches for an AGN-like radio core using VLBI techniques in the
Medusa have so far not detected any bright compact radio
components (Lonsdale, Lonsdale \& Smith 1992; van~Breugel et al. 1981).

In addition to this continuum information these spectral line observations,
along with those of AH00 provide information regarding the gas
velocity distribution within the central region of NGC\,4194 which can
be used to place some limits on the mass enclosed
within the central few arcsec. From these H{\sc i}
observations we observe a velocity gradient in the neutral gas of
$\sim$320\,km\,s$^{-1}$kpc$^{-1}$ over the central 2\farcs5
(470\,pc) which, if assumed to be in circular motion, implies a limit
on the enclosed mass of $\sim$10$^6$\,M$_{\odot}$. Such a limit constrains the mass of any black hole within the centre of the NGC\,4194.

However, considering the limits on the luminosities and sizes of these
 radio continuum components, the presence or not, of a weak AGN located at the position
of one of these two components cannot be categorically ascertained. The evidence
from the VLBI studies of Lonsdale et al. (1992) show that no high
brightness temperature synchrotron $>$1.8\,mJy at $\lambda=$18\,cm  is
detected. This supports the case for little or no AGN related radio
emission. Although it should be noted that 
southern nucleus of the binary AGN NGC\,6240 contains a very weak
compact VLBA core with flux of $\approx$1\,mJy (Gallimore \& Beswick
2004) which is free-free absorbed at 18\,cm. Further high resolution
multi-frequency radio imaging and X-ray observations are required to
investigate the possible scenario that there is an AGN buried in the
centre of this source.

\subsubsection{The starburst \& star-formation rate}

As has already been stated the Medusa merger is dominated, in terms of
its optical and infrared emission by starburst activity.  Starburst
emission results in relativistic charged particles accelerating in
supernovae remnants (SNRs) and radio supernovae (RSNe), which produces
the radio emission, such as found in other nearby starburst galaxies
(e.g. M82 -- Muxlow et al. 1994). 
 
It has been shown, for example by Condon (1992)
and Cram et al. (1998), that the 1.4\,GHz radio emission from star
forming galaxies can be considered as a direct tracer for the
supernovae rate. If this is assumed, the star-formation rate (SFR) of
the source can also be estimated, albeit with an initial mass function
(IMF) adjusted to produce massive enough stars to form supernovae that
result in significant radio emission ({\sl i.e.} M$\ge$8\,M$_{\odot}$)
such that,

\begin{equation}
\mathrm{SFR_{1.4}}=5.5\times\frac{\mathrm{L_{1.4}}} {4.6\times10^{21}\mathrm{WHz^{-1}}}\mathrm{M_\odot yr^{-1}}
\end{equation}

\noindent (e.g. Haarsma et al. 2000; Hopkins, Schulte-Ladbeck \&
Drozdovsky 2002). In this equation the multiplying factor of 5.5 is
included to convert the mass range 5--100\,M$_\odot$ used by Condon
(1992) to the range 0.1--100\,M$_\odot$, assuming a Salpeter
IMF. The low angular resolution (arcsecond), 1.4\,GHz
VLA flux of NGC\,4194 is 92.7\,mJy (Condon et al. 1990),  implies a total SFR$_{1.4}\approx$20\,M$_{\odot}$\,yr$^{-1}$.  In the case of
the Medusa, and most starburst and ULIRG systems, the accuracy of the
SFR as derived from optical data (e.g. H$\alpha$ observations) will be
severely affected by the large and variable nature of the obscuration
in this source. Using H$\alpha$ observations Storchi-Bergmann,
Calzetti \& Kinney (1994) quote a SFR$_{\rm
H\alpha}\approx$38\,M$_\odot$\,yr$^{-1}$ (assuming
H$_0$=50\,km\,s$^{-1}$\,Mpc$^{-1}$) which corresponds to a more modest
value of $\sim$17\,M$_\odot$\,yr$^{-1}$, if
H$_0$=75\,km\,s$^{-1}$Mpc$^{-1}$ is assumed. This value shows a
remarkable consistency with the SFR$_{1.4}$ value obtained from the
total radio flux considering the uncertainties that are introduced by
the high levels of nuclear obscuration within NGC\,4194.
    
In our higher resolution radio data only the emission from the
intense, compact nuclear star-forming regions can be
detected. Applying equation\,1 to our total MERLIN flux of 28\,mJy
implies a SFR of $\sim$6\,M$_\odot$\,yr$^{-1}$ for the
Medusa;  compared to the
far-infrared derived SFR$_{fir}$=6--7\,M$_\odot$\,yr$^{-1}$ (AH00). Clearly in 
the case of this radio determination the value can only be considered as a
lower limit for the total SFR, since it only accounts for
star-formation in the compact nuclear region. 

Compact clumps of star-formation have been imaged in NGC\,4194 by
Weistrop et al. (2004) at UV wavelengths using the {\sl HST}. They
find that there are $\sim$40 knots within the centre of
NGC\,4194. From their analysis they conclude that the total SFR
within the youngest ($<$20Myr) of these knots, which are primarily
located in the nuclear region in areas broadly consistent with the
distribution of the 1.4\,GHz MERLIN radio continuum, is
$\sim$6\,M$_\odot$\,yr$^{-1}$. This is consistent with our radio and
far-infrared determinations of the SFR in the nuclear region.   

The structure and luminosity of the radio emission in the lower
resolution 1.4\,GHz radio images (Fig.\,3) can easily be reproduced
via pure starburst mechanisms with only a relatively modest nuclear SFR. In this
scenario there is no requirement for a compact radio AGN although it
is not precluded.

\subsection{The ISM in the nucleus of the Medusa}
\subsubsection{Distribution of H{\sc i}, CO and dust}

Using these absorption data only gas in front of the radio
continuum can be detected. However it is also apparent the H{\sc i} distribution throughout the nuclear
region is not an even screen of gas. There are
significant variation in the measured opacity of the absorbing H{\sc
i} (assuming a constant T$_{\rm spin}$\footnote{Variations in the
value of T$_{\rm spin}$ across the radio continuum will also effect
the line strength (see Sec. 4.4.2).}), with
high columns of gas situated toward the southern part of the radio
emission. This area of increased H{\sc i} column density is spatially
coincident with the increased optical extinction from dust that arises
from the cross-nuclear dust lane (see Fig.\,1).  The implication of
this is that, in addition to the optical emission, the radio emission
is situated behind the dust lane and that it is probable that the
obscuring dust and neutral gas are physically associated. This is commonly seen in many nearby active galaxies with
nuclear dust lanes ({\sl e.g.} Cole et al. 1998; Beswick et al. 2002,
2003, 2004; Jackson et al. 2003).  

If we assume a standard Galactic gas-to-dust ratio (Stavely-Smith \&
Davies 1987), an inferred optical extinction can be derived from the
H{\sc i} column density. Assuming these ratios the optical extinction
at B-band is  A$_{\rm B}=8.62\times10^{-22}$N$_{\rm
H}$\,mag. Consequently a column density of
$\sim$5$\times$10$^{21}\,{\frac{\rm T_{spin}}{100\,{\rm
K}}}$\,atoms\,cm$^{-2}$ implies an extinction along the same line of
sight as the dust lane (positions 5, 6
and 7), of A$_{\rm B}\approx$4--5\,mag, assuming constant T$_{\rm
spin}\approx$100\,K. This value is approximately 50 times larger than
the implied levels of extinction against the compact radio components
toward the north of this radio source which are not crossed by the nuclear
dust lane (see Fig.\,1). 

AH00 mapped the distribution of CO (J$=$1$\rightarrow$0) in the
nucleus of NGC\,4194 at high angular resolution using OVRO. In this
study the CO emission in the velocity range
2506$\rightarrow$2570\,km\,s$^{-1}$ (see figure\,7 of AH00) is also found to trace the primary
dust lane of the Medusa.  AH00 observed CO emission to have linewidths of
FWHM $\sim$150\,km\,s$^{-1}$, larger but still comparable to those observed here (see
Table.\,2). This is quite remarkable considering
the angular resolution of these H{\sc i} observations is an order of
magnitude higher than the CO emission measurements. The consistency
of these two linewidth measurements could imply that these two
 gas components, situated along the same line of sight, are well mixed
over a range of angular scales ($\sim$2$\rightarrow$0.2 arcsec). 

At the areas of strongest CO emission
({\bf a} and {\bf b} in Fig.\,1 and AH00) the CO gas surface density is
$\sim$500--1000\,M$_{\odot}$\,pc$^{-2}$
($\sim$4--8$\times10^{21}$\,molecules\,cm$^{-2}$). These columns
compare to the H{\sc i} absorbing column densities of
$\sim$6--12$\times10^{21}$\,atoms\,cm$^{-2}$, assuming T$_{\rm
spin}=$100\,K.  These H{\sc i} columns
are merely a lower limit since they are only sensitive to the gas in
front of the radio continuum.  As stated above these observations also
show an increased column density of H{\sc i} toward the radio
continuum situated along the line of sight of the dust lane, which is
also coincident with the CO, in particular CO emission peaks {\bf a} (AH00 and Figs.\,1 \& 4).
Thus it implies that the nuclear CO, H{\sc i} and dust are all
situated along the same sight line.

\begin{figure}
   \centering
   \includegraphics[angle=0,width=8cm]{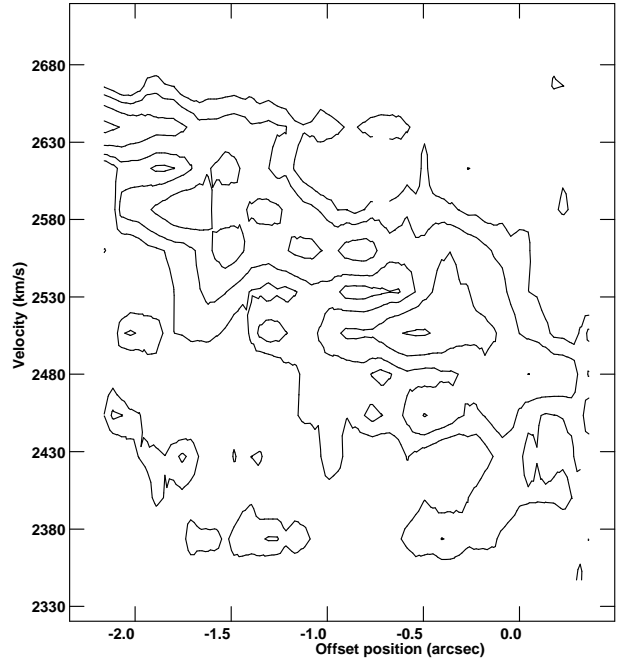}
      \caption{A position velocity diagram of H{\sc i} absorption in
the central region of the Medusa merger. The absorption has been
averaged over the entire R.A. range of the radio source, where the
radio continuum is in excess of 0.6\,mJy\,beam$^{-1}$. The contour
levels are set at $-$8, $-$5.657, $-$4, $-$2.282, $-$2, $-$1.414, $-$1
and 1 times 0.36\,mJy\,beam$^{-1}$. The position axis is declination
relative to 54\degr48\arcmin16\farcs4.}
            \label{Fig5}
\end{figure}

\subsubsection{Dynamics of the cold ISM: the fuel for the
evolving starburst}

Figures\,4{\bf c} and 5 show the velocity field and a north-south
position-velocity diagram derived from these H{\sc i} absorption
observations. Comparing the sub-arcsec angular resolution H{\sc i}
absorption velocity field (Fig.\,4{\bf c}) with the centre of the
$\sim$1\farcs7 angular resolution CO velocity field presented in
Fig.\,8{\bf b} of AH00 it is apparent that the `global' nuclear
CO and H{\sc i} dynamics are consistent. In both of these cases the
general trend is for the gas to be rotating following a north-south
velocity gradient of a few tens of km\,s$^{-1}$ per arcsec; further confirming
the likely association of these two components. 

On the scale of the OVRO CO emission the radio continuum peak
is coincident with the kinematic centre of the galaxy. Centred close
to this kinematic centre, these MERLIN observations provide a 'close-up' view of the nuclear gas
dynamics. In Fig.\,4{\bf c} the velocity field, although consistent
with the CO, shows significant deviations and smaller scale velocity
structure that is smoothed away by the larger CO synthesised
beam. In particular the isovelocity contours of Fig.\,4{\bf c} show a
significant twist toward the west as they
cross the area of peak radio flux density (see also spectra 3 and 4
of Fig.\,3). This twist in the velocity field is hinted at in the CO
observations of AH00. However AH00 conclude that the gas in this
region undergoes approximately solid-body rotation. This is broadly
confirmed by these results, albeit only for a small cross-section
through the rotating material, but it is apparent that the gas does
 also possess some additional non-circular motions superimposed upon this rotation.

As discussed in the previous section the H{\sc i}
absorption shows significant inhomogeneities in its column densities. In particular it is clear that the gas column densities are
significantly lower against the bright radio continuum peak in the
north of the source; along lines of sight where the gas is
blueshifted relative to the galaxy's systemic velocity. Of course
these column density variations 
may, in part, be the result of an underlying gas temperature gradient.
However such an effect would need to be extreme to cause the large
observed opacity differences and should not considered to be a
dominant factor. Additionally AH00 find that the distribution of CO
emitting gas is also asymmetric about the galaxy's kinematic centre. They observe
bright, compact CO emission components redshifted relative to the
systemic velocity of the galaxy and offset from the radio core by
$\sim1\arcsec$; no equivalent emission features are observed at similar
blueshifted velocities.

Assuming that the radio continuum emission is related to the amount of recent star-formation
occurring at anyone point; some of the areas
with the most active ongoing star-formation will be located toward the
brightest radio components. Further evidence from UV {\it
HST} observations by Weistrop et al. (2004), where high concentration of young
($\le$20\,Myr old) star clusters are imaged, confirms that the majority
of the recent star-formation activity is situated toward the north and
western parts of NGC\,4194. Thus it appears that a
considerable portion of the activity is situated on the blueshifted side of the source and away from the bulk
of the cold gas. 

Unlike emission experiments these absorption observations only detect gas
that is in front of the  source. Therefore if the star-formation
areas to the north and west are being
fed by the cold gas it is possible that the H{\sc i}
absorption column density toward the radio continuum
peaks is depleted. Dynamically this might be considered as following the observed
rotation velocity, where the gas approaches the rear of this area of
star-forming region, is partially converted to stars during its
passage and hence has a reduced column density in front of this region.  This scenario implies
that the primary reservoir of gas that is fuelling these regions of
starburst activity is in fact the gas toward the south of the radio
continuum, where both the column densities of H{\sc i} absorption and
CO emission peak (i.e. an area along the same line-of-sight as the
nuclear dust lane). Of course in this hypothesis it is equally
plausible that the
site of the active star-formation is moving toward
the region with high gas density as the starburst evolves rather than
the gas itself being transported into the stationary star-formation region.

\subsubsection{Physics of the cold gas}

As stated in Sec.\,3.2 the column density of neutral hydrogen, N$_{\rm
H}$, is related to the gas spin (excitation) temperature, T$_{\rm
spin}$ and the observed H{\sc i} absorption by

\begin{equation}
{\rm N}_{\rm H}=1.823\times 10^{18}{\rm T_{spin}}\int{\tau}{\rm dV}\,\,{\rm atoms\,cm^{-2}}
\end{equation}

\noindent where V is the in units of km\,s$^{-1}$ and $\tau$ is the
measured opacity of the line. Thus any calculated values of  N$_{\rm
H}$ will be dependent upon the excitation temperature, T$_{\rm spin}$. 

The value of T$_{\rm spin}$ for neutral hydrogen is controlled by three 
factors; pumping by Ly$\alpha$ photons, the absorption of 21\,cm continuum 
radiation, and collisions (Field 1959a, 1959b, 1959c). It is well known 
that in high density regions that collisional excitation methods dominate; 
thus in these regions the spin temperature of H{\sc i} will approach the 
gas kinetic temperature. For example measurements of H{\sc i} emission and 
absorption column densities through dense regions of galactic disks are 
often consistent for excitation temperature in the range of a few 10s to a 
few 100\,K (e.g. Payne, Salpeter \& Terzian 1983; Liszt et al. 1983; Braun 1997; Wolfire et al. 2003 and refs therein).  It should be noted 
here that these calculations assume that the atomic gas, observed by 
either emission or absorption, is in a single component medium in 
equilibrium and thermalized (i.e. n$\gtsim$1000\,cm$^{-3}$); see for 
example Kulkarni \& Heiles (1988). The assumption that T$_{\rm spin}$ can 
be approximated to the gas kinetic temperature, T$_{k}$, however, does not 
necessarily hold true especially (if a two component medium is assumed [a 
cold and warm neutral medium]) for areas of lower density gas clouds that 
are subject to irradiation by either significant levels Ly$\alpha$ photons 
or 21\,cm continuum radiation (Wolfire et al. 2003). Work by Gallimore et 
al. (1999) further investigated these excitation processes addressing 
H{\sc i} absorbing gas clouds within a few tens of parsecs of Seyfert 
nuclei. One of the primary conclusions of this work was that even in these 
regions where both an abundant 21\,cm and Ly$\alpha$ source is present, 
T$_{\rm spin}$ will asymptotically approach T$_{k}$ for gas densities of 
n$\gtsim$1000\,cm$^{-3}$ (i.e. a collision dominated regions). Whereas 
below this gas density the H{\sc i} spin temperature, if in the presence 
of a 21\,cm and/or Ly$\alpha$ photon source, will be significantly 
influenced by these other excitation processes resulting in T$_{\rm spin}$ 
deviating away from the kinetic gas temperature (see figure 12 of 
Gallimore et al 1999). However at this point we should also highlight the 
fact that for H{\sc i} absorption, $\tau\propto({\rm T_{spin}\Delta{\rm 
V})^{-1}}$. Thus H{\sc i} absorption studies, such as these, are most 
sensitive to the coldest atomic gas clouds.

Regarding the Medusa merger we have already assessed the plausibility of 
either an AGN embedded within the central starburst of this galaxy or 
whether the radio emission detected can be accounted for by star 
formation alone (see Sec.\,4.3). From this discussion, although the presence 
of a weak or highly obscured active core cannot be eliminated, we conclude 
that the dominant energy input into the ISM of the Medusa is via the 
efficient starburst at its centre. Thus this ongoing star formation will 
be the primary heating processes for the ISM of NGC\,4194.

The molecular ISM of the Medusa merger is characterized by an elevated 
CO/$^{13}$CO 1-0 molecular line ratio of $19 \pm 4$ and relatively faint 
HCN 1-0 emission ($^{12}$CO/HCN$>$25)\footnote{Recent observations by S. 
Aalto (unpublished) have detected very faint HCN emission from the 
compact, southern molecular peak which could be indicative of young or 
ongoing but obscured star formation in this region.}. The faintness of HCN 
emission suggests that the average density in the molecular medium is 
below 10$^4$\,cm$^{-3}$ and we suggest that the weak $^{13}$CO 1-0 line is 
caused by two effects: high gas temperatures (T$_k >$50 K) and the presence 
of diffuse, unbound gas. High resolution observations reveal that the 
$^{13}$CO 1-0 line emission is relatively weak primarily in parts of the 
central dust lane and towards the bright central CO peak (Aalto et al in 
prep). While the latter is likely caused by elevated kinetic temperatures, 
the former is probably related to diffuse gas in a density-wave like 
situation, similar to that of the bar of NGC~7479 (H\"uttemeister et al 
2000). In general most galaxies with elevated values of the CO/$^{13}$CO 
1-0 line ratio also have $f(60\mu{\rm m}/f(100 \mu {\rm m})\gtsim0.8$ 
indicating a high average dust temperature (Aalto et al. 1991, 1995). This 
further supports the notion that gas and dust is being heated by the 
ongoing starburst.

Considering these arguments it can be concluded that the molecular gas
within the centre of the Medusa merger firstly exists in a relatively
low density environment. In this sense, although there will clearly be
higher density molecular gas clouds, the majority of the gas is in
regions with n$\sim$1--5$\times$10$^3$\,cm$^{-3}$. Additionally the molecular
component appears to have slightly elevated kinetic temperatures (T$_k\gtsim$50\,K). This
provides us with an insight into the physics of the molecular clouds
within the Medusa which can be translated into an upper limit for the
density of the atomic gas and a lower limit on its kinetic
temperature. These limits are valuable in that they help to define one
extreme of the cold atomic gas phase within this merger. The primary
caveat is that in most `cartoon' models of the ISM within galaxies
the, coldest and densest, molecular clouds are cocooned within an
envelope of more diffuse atomic gas which will be partially
ionised. Thus, apart from the unlikely situation where the mixing of
the atomic and molecular gas is complete, these molecular diagnostic
tool can only be used to infer limits for the physics of the atomic
gas phase.

\section{Conclusions}    

Using radio continuum and H{\sc i} absorption observations made with
MERLIN we have studied the neutral hydrogen gas distribution and
kinematics, in addition to the 1.4\,GHz radio continuum structure of the
central kiloparsec of the Medusa merger at sub-arcsecond angular
resolutions. These observations constitute the highest angular
resolution study, to date, of either the neutral or molecular gas
within this source.

Our highest resolution 1.4\,GHz images of NGC\,4194 reveal a pair of
compact mJy radio components situated at the position of the peak in
lower resolution radio images of this source. These two components are
surrounded, primarily toward the south by a more diffuse region of weaker radio emission. These compact radio components are
unresolved with sizes of $<$25\,pc
and a separation of $\sim$65\,pc. We have used these radio
observations to investigate the energy source of the radio emission
and compare the two most probable scenarios; firstly that some portion
of the radio emission is related to the presence of a weak AGN
component in addition to the well known starburst emission from this
source, or secondly, whether all of the radio emission can completely
account for the extended nuclear starburst. Using a combination of
the H{\sc i} gas dynamics and star-formation rates, derived from the
nuclear radio emission compared, to those derived from observations at other
wavelengths, we have tentatively concluded that the majority, if not
all, of the emission is related to star-formation activity. At present
we cannot categorically eliminate the possibility of a weak AGN
component without further observational studies such as further high
sensitivity multi-frequency VLBI observation or high resolution X-ray
observations.

By observing H{\sc i} via absorption we have been able to trace
the gas dynamics in front of the background radio continuum within the
half central kiloparsec of the Medusa merger. The distribution of the
absorbing H{\sc i} gas shows a significant enhancements toward the
south of the radio continuum source. This area is along the same
line-of-sight as the peaks in the $^{12}$CO emission observed by AH00
and is also co-spatial with a cross-nuclear dust lane (see
Fig.\,1). Following the results of AH00 we conclude that all three of
these cold ISM components (dust, CO and atomic hydrogen) are probably
physically related, with both the CO emission and H{\sc i} absorption
exhibiting similar dynamics. The H{\sc i} absorption traces a
relatively shallow, approximately, north-south velocity gradient of
$\sim$320\,km\,s$^{-1}$\,kpc$^{-1}$ over $\sim\frac{1}{2}$\,kpc of the
background radio continuum. This gradient is consistent with that
observed in CO by AH00. Considering this gradient and the distribution
of the absorbing H{\sc i} and the most active star-forming regions we
discuss the possible role of the gas dynamics with respect to the
fuelling of the starburst. Since the Medusa merger has been
extensively studied via a variety of molecular gas transitions we have
also discussed the use of these as diagnostic tools to infer some
limits on the physical environment that the molecular and atomic gas
resides in.

\begin{acknowledgements}

RJB would like to acknowledge financial support by the European
Commission's I3 Programme ``RADIONET'' under contract No.505818. We would like to thank all
the MERLIN staff for their assistance with these observations.  In
addition we thank Susan Neff for kindly sharing with us her recent VLA
observations of NGC\,4194 prior to publication.  MERLIN is a national
facility operated by the University of Manchester on behalf of PPARC
in the UK. The authors wish to thank the Canadian Astronomy Data
Centre, which is operated by the Hertzberg Institute of Astrophysics,
National Research Council of Canada for providing recalibrated {\sl
HST} data to the author as a guest user.

\end{acknowledgements}

\end{document}